\documentclass[11pt]{article}
\usepackage[utf8]{inputenc}
\usepackage[left=1in,right=1in]{geometry}
\usepackage{authblk}
\usepackage[pdfstartview=FitH,pdfpagemode=None]{hyperref}
\usepackage{amsmath}
\usepackage{amsfonts}
\usepackage{slashed}
\usepackage{cite}
\title{Conformal Primary Basis for Dirac Spinors}
\author[1]{Lorenzo Iacobacci \thanks{E-mail: \texttt{lorenzoiacobacci@gmail.com}}}
\author[1,2]{Wolfgang Mück \thanks{E-mail: \texttt{mueck@na.infn.it}}}
\affil[1]{Dipartimento di Fisica ``Ettore Pancini'', Universit\`a degli Studi di Napoli ``Federico II'' \authorcr Via Cintia, 80126 Napoli, Italy}
\affil[2]{Istituto Nazionale di Fisica Nucleare, Sezione di Napoli \authorcr Via Cintia, 80126 Napoli, Italy}
\date{\today}
\newcommand{\ie}{i.e.,\ }

\newcommand{\rmd}{\,\mathrm{d}}

\newcommand{\re}{\operatorname{Re}}

\newcommand{\e}[1]{\operatorname{e}^{#1}}

\newcommand{\vo}{\vec{w}}
\newcommand{\vz}{\vec{z}}
\newcommand{\vk}{\vec{k}}

\newcommand{\slo}{\slashed{w}}
\newcommand{\slz}{\slashed{z}}
\newcommand{\slk}{\slashed{k}}

\newcommand{\hp}{\hat{p}}

\newcommand{\BesselK}{\operatorname{K}}
\newcommand{\BesselJ}{\operatorname{J}}

\numberwithin{equation}{section}
\begin{document}
\maketitle
\begin{abstract}
We study solutions to the Dirac equation in Minkowski space $\mathbb{R}^{1,d+1}$ that transform as $d$-dimensional conformal primary spinors under the Lorentz group $SO(1,d+1)$. Such solutions are parameterized by a point in $\mathbb{R}^d$ and a conformal dimension $\Delta$. The set of wavefunctions that belong to the principal continuous series, $\Delta =\frac{d}2 + i\nu$, with $\nu\geq 0$ and $\nu \in \mathbb{R}$ in the massive and massless cases, respectively, form a complete basis of delta-function normalizable solutions of the Dirac equation. In the massless case, the conformal primary wavefunctions are related to the wavefunctions in momentum space by a Mellin transform. 
\end{abstract}

%%%%%%%%%%%%%%%%%%%%%%%%%%%%%%%%%%%%%%%%%%%%
%
% structure
%
% 1) intro
% 2) embedding space formalism
% 3) conformal primary basis for massive spinor fields
% 4) conformal primary basis for massless spinor fields
% 5) conclusions
% appendices:
%   A) recurrent formulas and integrals
%	B) Dirac inner product
%	C) Shadow transform
%	D) completeness
%
%%%%%%%%%%%%%%%%%%%%%%%%%%%%%%%%%%%%%%%%%%%%

\section{Introduction and Summary}
\label{intro}

The classification of fields according to the irreducible representations of a symmetry group has been, since the work of Wigner \cite{Wigner:1931}, an essential ingredient in Quantum Mechanics and Quantum Field Theory. In the case of fields in Minkowski space-time, which enjoys symmetry under the Poincar\'e group, Wigner's 1939 paper \cite{Wigner:1939cj} still provides the foundation of our description of particles of any spin. This description is based on wavefunctions in momentum space, where translation symmetry is manifest. In four space-time dimensions, an alternative approach is to work in twistor space \cite{Penrose:1967wn, Witten:2003nn} which gives, for example, an handle on MHV amplitudes \cite{Parke:1986gb} and the BCFW recursion relations \cite{Britto:2005fq}.
Recent years have seen a surge of activity on flat space holography \cite{deBoer:2003vf} focussing on the imprint of two-dimensional conformal symmetry in four-dimensional scattering amplitudes. In this context, conformal primary wavefunctions have been constructed \cite{deBoer:2003vf, Cheung:2016iub, Pasterski:2016qvg, Pasterski:2017kqt}, which has allowed to formulate scattering amplitudes as conformal correlators on the celestial sphere and study their properties in the conformal basis \cite{Pasterski:2017ylz, Ball:2019atb, Pate:2019lpp, Fotopoulos:2019tpe, Law:2019glh, Law:2020tsg, Fotopoulos:2020bqj, Law:2020xcf}. In particular, the construction of conformal primary wavefunctions for massless and massive scalars, photons and gravitons, in any dimension, and the proof of their completeness was carried out by Pasterski and Shao \cite{Pasterski:2017kqt}, with the subtle extension of the soft modes for photons and gravitons added in \cite{Donnay:2018neh}. 

In this paper, we generalize the work of Pasterski and Shao to Dirac spinors.\footnote{Massless spin-$\frac12$ fermions in four dimensions have been considered in \cite{Fotopoulos:2020bqj} in the context of the celestial CFT of supergravity/YM theory. We would also like to mention \cite{Narayanan:2020amh}, which appeared shortly after our manuscript was posted on arXiv.} This is motivated partly by possible applications of flat space holography in QED and for the explicit construction of the conformal partial wave decomposition \cite{Dobrev:1975ru} in CFT. We construct solutions of the massive and massless Dirac equation in Minkowski space-time $\mathbb{R}^{1,d+1}$ that transform as conformal primary spinors in Euclidean $\mathbb{R}^d$, calculate their inner product and their shadow transform \cite{Ferrara:1972uq, Ferrara:1972xe, Ferrara:1972xq, Ferrara:1973vz}, and prove their completeness. As one may expect from a possible supersymmetric extension, our results retrace the case of scalar fields. For massive Dirac spinors, the basis wavefunctions are parameterized by a vector $\vo \in \mathbb{R}^d$ representing a boundary point in the momentum space of the particle, and a conformal dimension $\Delta$, which belongs to one half of the principal continuous series, 
\begin{equation}
\label{intro:basis1}
	\Delta \in \frac{d}2 + i\mathbb{R}_{\geq 0}\qquad (m>0)~.
\end{equation}
Alternatively, one may use their shadow transforms, which lie on the other half.
In the massless case, which will be obtained by considering the $m\to 0$ limit, the conformal dimensions of the basis wavefunctions span the entire principal continuous series, 
\begin{equation}
\label{intro:basis2}
	\Delta \in \frac{d}2 + i\mathbb{R} \qquad (m=0)~.
\end{equation}
In both cases, the wavefunctions also carry a $d$-dimensional Euclidean spinor index parameter ($\alpha=1,2,\ldots, n$, with  $n=2^{[\frac{d}2]}$), which will, however, be left implicit throughout the paper.

Our construction is based on the embedding space formalism \cite{Dirac:1936fq, Mack:1969rr, Weinberg:2010fx, Costa:2011mg}, which exploits the isomorphisms between the Lorentz group in $\mathbb{R}^{1,d+1}$, the symmetry group of the hyperbolic space $\mathbb{H}_{d+1}$ (Euclidean anti-de Sitter space) and the conformal group in $d$-dimensional Euclidean space $\mathbb{R}^d$. Here, $\mathbb{R}^{1,d+1}$ is to be considered as momentum space, $\mathbb{H}_{d+1}$ as the space of on-shell momenta for massive particles, and $\mathbb{R}^d$ parameterizes null momenta. The embedding space formalism is a powerful tool in the conformal bootstrap programme \cite{Ferrara:1973yt, Polyakov:1974gs}, for some recent applications see \cite{Costa:2011dw, SimmonsDuffin:2012uy, Iliesiu:2015akf, Iliesiu:2015qra, Isono:2017grm, Fortin:2019dnq}. In particular, in \cite{Fortin:2019dnq} the uplift to embedding space has been constructed for conformal primary fields of any spin. However, our treatment of spinors will follow \cite{Isono:2017grm}. In particular, we will consider Dirac spinors, which exist in any dimension. The reduction to Weyl or Majorana spinors, when they exist, is straightforward. 

The rest of the paper is structured as follows. In Sec.~\ref{embedding}, we review the embedding space formalism. The massive and massless conformal primary spinor wavefunctions are discussed in Sections~\ref{massive.case} and \ref{massless.case}, respectively. These sections are kept reasonably concise, with all long calculations included in the appendices.

\section{Embedding Space}
\label{embedding}

Consider Minkowski space $\mathbb{R}^{1,d+1}$, with coordinates $P^M$, $M=0,1,\ldots, d+1$. In light-cone coordinates, which we use throughout this paper, we represent $P^M$ by
\begin{equation}
\label{emb:lightcone}
	P^M = \left(P^+, P^-, \vec{P}\right)^T~, 
\end{equation}
with $P^\pm = P^0 \pm P^{d+1}$. The non-zero components of the Minkowski metric, $\eta_{MN}$, in light-cone coordinates are\footnote{This corresponds to  $(-++\cdots+)$ signature.}
\begin{equation}
\label{emb:Mink.metric}
	\eta_{+-}=\eta_{-+}=-\frac12~,\qquad \eta_{ij}=\delta_{ij}\qquad (i, j= 1,2,\ldots, d)~. 
\end{equation}

The $(d+1)$-dimensional hyperbolic space $\mathbb{H}_{d+1}$ can be embedded in $\mathbb{R}^{1,d+1}$ by restricting to time-like unit vectors $\hp^M$ ($\hp^2 = -1$) with $\hp^+>0$. The components $\hp^\mu$ ($\mu=1,2,\ldots, d+1$) provide a good set of coordinates on $\mathbb{H}_{d+1}$, but it is more useful to represent $\hp\in \mathbb{H}_{d+1}$ by 
\begin{equation}
\label{emb:H.coord}
	\hp^M = \frac1{y} \left(1, y^2 +|\vz|^2, \vz\right)^T~.
\end{equation}
In the coordinates $(y, \vz)$, the induced metric on $\mathbb{H}_{d+1}$ is
\begin{equation}
\label{emb:H.metric}
	\rmd s^2_{\mathbb{H}_{d+1}} = \frac1{y^2} \left( \rmd y^2 + \rmd \vz^2\right)~.
\end{equation}
The invariant volume measure for integration over $\mathbb{H}_{d+1}$ is given by
\begin{equation}
\label{emb:H.integ}
	[\rmd \hp] = \frac{\rmd^{d+1} \hp}{\hp^0} = \frac{\rmd y \rmd^d z}{y^{d+1}}~.
\end{equation}

The $d$-dimensional Euclidean space $\mathbb{R}^d$ can be embedded in $\mathbb{R}^{1,d+1}$ as the projective null cone, defined by null vectors $Q^M$ ($Q^M Q_M =0$), with $Q^+>0$ and identifying $Q^M\sim \lambda Q^M$ for any $\lambda>0$. Writing 
\begin{equation}
\label{emb:Q.def}
	Q^M = Q^+ q^M(\vo)~, \qquad q^M(\vo) = \left( 1, \vo^2 , \vo \right)^T~,
\end{equation}
$q^M(\vo)$ is a representative on the projective null cone, and the $w^i$ ($i=1,2,\ldots , d$) are the coordinates on $\mathbb{R}^d$, with the standard Euclidean metric.

Lorentz transformations on $\mathbb{R}^{1,d+1}$, which act linearly on $\hp^M$ and $Q^M$, 
\begin{equation}
\label{emb:Lor.X}
	\hp'{}^M = \Lambda^M{}_N\, \hp^N~,\qquad  Q'{}^M = \Lambda^M{}_N\, Q^N~,
\end{equation}
act non-linearly on the coordinates of $\mathbb{H}_{d+1}$ and $\mathbb{R}^d$. The transformed coordinates $(y', \vz')$ and $\vo'$ follow from \eqref{emb:Lor.X} via \eqref{emb:H.coord} and \eqref{emb:Q.def}, respectively. For $\mathbb{R}^d$, the \emph{projective} property of the embedding leads to  
\begin{equation}
\label{emb:q.trafo}
	\Lambda^M{}_N\, q^N(\vo) = \left\Vert \frac{\partial \vo'}{\partial \vo} \right\Vert^{-\frac1{d}} q^M(\vo')~,  
\end{equation}
where the transformation $\vo \to \vo'$ is a \emph{conformal} transformation, and $\left\Vert \frac{\partial \vo'}{\partial \vo} \right\Vert$ is its Jacobian determinant. We refrain from repeating the explicit expressions for the coordinates  $(y', \vz')$ and $\vo'$ in the two cases, as they can be found elsewhere \cite{Pasterski:2017kqt} and will not be needed.\footnote{In addition to the transformations listed in \cite{Pasterski:2017kqt}, one may consider the discrete transformation that swaps $X^+$ with $X^-$ (reflection of the $(d+1)$-th direction). This acts as an inversion on $\vo$.}

The construction of conformal primary wave functions for massive fields rests upon the transformation laws \eqref{emb:Lor.X} and \eqref{emb:q.trafo}. Indeed, these two equations imply that the Lorentz scalar 
\begin{equation}
\label{emb:scalar}
	-2 \hp \cdot q(\vo) = \frac{y^2 +|\vz -\vo|^2}{y} 
\end{equation}
has weight $-1$ under the induced conformal transformation of $\vo$.

Let us now introduce the Dirac gamma matrices that we need for our treatment of (Dirac) spinors. 
Let $\gamma_i$ be $d$-dimensional Euclidean gamma matrices satisfying the Clifford algebra 
\begin{equation}
\label{emb:Clifford.Euclidean}
	\gamma_i \gamma_j + \gamma_j \gamma_i = 2 \delta_{ij}~.
\end{equation}
These are $n \times n$ square matrices, where $n=2^{\left[\frac{d}2\right]}$. Throughout the paper, the standard slashed shorthand will be used only for $d$-dimensional quantities, such as
\begin{equation}
\label{emb:shorthand}
	\slo = \gamma_i w^i~.
\end{equation} 

The Dirac gamma matrices $\Gamma_M$ for the embedding space $\mathbb{R}^{1,d+1}$ will be taken as
\begin{equation}
\label{emb:gamma.mat}
	\Gamma_i = 
	\begin{pmatrix}
		\gamma_i & 0 \\
		0 & -\gamma_i
	\end{pmatrix}~,\qquad 
	\Gamma_- =  
	\begin{pmatrix}
		0 & -1 \\
		0 & 0
	\end{pmatrix}~,\qquad 
	\Gamma_+ =  
	\begin{pmatrix}
		0 & 0 \\
		1 & 0
	\end{pmatrix}~,
\end{equation} 
where each entry denotes an $n \times n$ block.\footnote{For even $d$, these matrices give rise to a reducible representation of the Lorentz group. The embedding formalism of irreducible (Weyl) representations in even dimensions has been considered in \cite{Fortin:2019dnq}, but uses a different convention for the gamma matrices.} They satisfy
\begin{equation}
\label{emb:Clifford:Mink}
	\Gamma_M \Gamma_N + \Gamma_N \Gamma_M = 2 \eta_{MN}~.
\end{equation}
We remark that $\Gamma_+ + \Gamma_- = \Gamma_0 = - \Gamma^0$, which is needed in some calculations.

The embedding formalism for (Dirac) spinors in general dimensions has been developed in \cite{Isono:2017grm}. With our conventions for the gamma matrices, a Dirac spinor is written in the block form 
\begin{equation}
\label{emb:spinor.decomp}
	\Psi = 
	\begin{pmatrix}
		\Psi_+ \\ \Psi_-	
	\end{pmatrix}~.
\end{equation}
Clearly, $\Psi_\pm$ transform as Dirac spinors in $\mathbb{R}^d$. If one assumes further that, on the light cone, $\Psi_\pm$ are homogeneous functions of $Q$ of degree $-(\Delta+\frac12)$, \ie
\begin{equation}
\label{emb:hom.spin}
	\Psi(\lambda Q) = \lambda^{-(\Delta +\frac12)} \Psi(Q)~,
\end{equation}
and if one defines 
\begin{equation}
\label{emb:spinor.null.cone}
	\psi_\pm(\vo) = \Psi_\pm (q(\vo)) = (Q^+)^{\Delta +\frac12} \Psi_\pm(Q)~,
\end{equation}
then the combination 
\begin{equation}
\label{emb:spinor.conf}
	\psi(\vo) = \psi_+(\vo) - \slo \psi_-(\vo) 
\end{equation}
is also a Dirac spinor on $\mathbb{R}^d$ and transforms as a conformal primary of dimension $\Delta$ under the conformal transformations of $\vo$ that are induced by the Lorentz transformations of $Q$ \cite{Isono:2017grm}. In order to construct a conformal primary conjugate Dirac spinor on $\mathbb{R}^d$, which we will need in the next section, we manipulate \eqref{emb:spinor.conf} as follows,\footnote{In $\mathbb{R}^d$, conjugate simply means the hermitean conjugate, $\bar{\psi}= \psi^\dagger$, whereas $\bar{\Psi}= \Psi^\dagger \Gamma^0$ in $\mathbb{R}^{1,d+1}$.}
\begin{align}
\notag 
	\bar{\psi}(\vo) &= \bar{\psi}_+(\vo) - \bar{\psi}_-(\vo) \slo \\
\notag &= 
	\begin{pmatrix}
		\bar{\psi}_+(\vo) & \bar{\psi}_-(\vo)	
	\end{pmatrix}
	\begin{pmatrix}
		0 & 1 \\ -1 & 0
	\end{pmatrix}
	\begin{pmatrix}
		\slo \\ 1
	\end{pmatrix}\\
\label{emb:cong.spinor.conf}
	&= \bar{\Psi}(q(\vo))
	\begin{pmatrix}
		\slo \\ 1
	\end{pmatrix}~.
\end{align}

We avoid introducing spinors on $\mathbb{H}_{d+1}$, because this would imply an unnecessary distiction between even and odd $d$. 
%  notation and conventions, conformal symmetry of celestial sphere ...

\section{Conformal primary basis for massive spinor fields}
\label{massive.case}

\subsection{Solutions of the massive Dirac equation}
We are interested in solutions of the massive Dirac equation in $(d+2)$-dimensional Minkowski space,
\begin{equation}
\label{Dirac}
	\left( \Gamma^M\frac{\partial}{\partial X^M} - m \right) \Psi(X) = 0~,
\end{equation}
which transform as conformal primary (conjugate) spinors under the conformal transformations induced on the celestial sphere by space-time Lorentz transformations. Using the embedding space formalism, such solutions can be written down straightforwardly. 
Up to normalization, they are given by\footnote{As mentioned in the introduction, we keep all spinor indices implicit throughout the paper. In fact, $\Psi^\pm_\Delta(X,\vo)$ carries a space-time spinor index ($1,2,\ldots, 2n$) on the left, and an $\mathbb{R}^d$ (conjugate) spinor index ($1,2,\ldots, n$) on the right.}
\begin{equation}
\label{Dirac.sol}
	\Psi^\pm_\Delta(X;\vo) = \int [\rmd \hp] \frac{\e{\pm i m \hp \cdot X}}{[-2 \hp\cdot q(\vo)]^{\Delta+\frac12}} 
		\Pi_\pm(\hp) \begin{pmatrix} \slo \\ 1 \end{pmatrix}~.
\end{equation}
Here, $\pm$, $\Delta$ and $\vo$ are parameters. The $\pm$ signs correspond to the solutions with positive and 
negative energy, respectively. The projectors
\begin{equation}
\label{Pi.pm}
	\Pi_\pm(\hp) = \frac12 \left( 1 \pm i \hp^M \Gamma_M \right)
\end{equation}
make \eqref{Dirac.sol} a solution of \eqref{Dirac}. Furthermore, the column matrix on the right and the denominator ensure that $\Psi^\pm_\Delta(X;\vo)$ behaves as a conformal primary conjugate spinor of dimension $\Delta$ under conformal transformations of $\vo$, as reviewed in the previous section. Readers familiar with AdS/CFT might recognize an harmonic function in $\mathbb{H}_{d+1}$, but the dimensionality of the spinor would not match for even $d$.   

To obtain explicit solutions, one might follow \cite{Pasterski:2017kqt} and consider the Dirac equation directly using the solution ansatz
\begin{equation}
\label{ansatz}
	\Psi_\Delta(X;\vo) = \frac{f_1(X^2) + f_2(X^2) \Gamma^M X_M}{[-2q(\vo)\cdot X]^{\Delta+\frac12}} 
	\begin{pmatrix} \slo \\ 1 \end{pmatrix}~.
\end{equation} 
Note that there cannot be a term with $\Gamma^M q_M(\vo)$ in the numerator, because 
\begin{equation}
\label{st:proj}
	\Gamma^M q_M(\vo)
	\begin{pmatrix}
		\slo \\ 1	
	\end{pmatrix} =
	\begin{pmatrix}
		\slo & -w^2 \\ 1 & -\slo	
	\end{pmatrix}
	\begin{pmatrix}
		\slo \\ 1	
	\end{pmatrix} = 0~.
\end{equation} 
The Dirac equation translates into two coupled, first order differential equations for $f_1$ and $f_2$, which can be easily solved. This procedure, however, has two disadvantages. First, one needs to recover an overall multiplicative constant in the end in order to match with \eqref{Dirac.sol}. Second and more importantly, the choice of the branch cut that distinguishes between the positive and negative energy solutions remains ambiguous. Therefore, it is better to perform the integral in \eqref{Dirac.sol}, which we will do in App.~\ref{app.explicit}. The result is
\begin{align}
\label{Dirac.sol.explicit}
	\Psi^\pm_\Delta(X;\vo) &= 
	\frac{(\pm i)^{-\Delta-\frac12} (2\pi)^{\frac{d}2} m^{-\frac{d}2}}{[-2q(\vo)\cdot X\mp i\epsilon]^{\Delta+\frac12}} \\
\notag &\quad \times\left( 1+\frac1m \Gamma^M \partial_M \right)
	\left[ \left(\sqrt{X^2}\right)^{\Delta-\frac{d-1}2} 
  \BesselK_{\Delta-\frac{d-1}2} \left(m\sqrt{X^2}\right) \right]
  \begin{pmatrix} \slo \\ 1 \end{pmatrix}~,
\end{align}
where $\BesselK_\alpha$ is a modified Bessel function.

\subsection{Dirac inner product}

The Dirac inner product is defined as 
\begin{equation}
\label{inner.prod}
	\left( \Psi, \Psi' \right) = \int \rmd \Sigma^M\, \bar{\Psi} \Gamma_M \Psi' = \int \rmd^{d+1} X\, \Psi^\dagger(X) \Psi'(X)~,
\end{equation}
where the second equality holds upon choosing an $X^0=\text{const.}$ hypersurface for integration. The somewhat lengthy calculation of \eqref{inner.prod} for two conformal primary spinor wave functions \eqref{Dirac.sol} is deferred to App.~\ref{app.dirac}. Integrability restricts the conformal dimension $\Delta$ to the principal continuous series,
\begin{equation}
\label{ip:p.series}
	\Delta = \frac{d}2 +i \nu \qquad (\nu \in \mathbb{R})~,
\end{equation}
and the result is
\begin{align}
\label{inner.prod.sol}
	 \left( \Psi^\pm_{\frac{d}2+i\nu}(\vo), \Psi^\pm_{\frac{d}2+i\nu'}(\vo') \right) 
	 &= \left(\frac{2\pi^2}m\right)^{d+1} 
	 	\left[ \frac{\Gamma(\frac12-i\nu) \Gamma(\frac12+i\nu)}{\Gamma(\frac{d+1}2 -i\nu) \Gamma(\frac{d+1}2 +i\nu)}
	 	\delta(\nu - \nu') \delta^d(\vo-\vo') \right.\\
\notag
	&\quad \left. \pm i \frac{\Gamma(\frac12-i\nu)}{\pi^{\frac{d}2} \Gamma(\frac{d+1}2 -i\nu)}\delta(\nu + \nu') 
	\frac{\slo - \slo'}{|\vo -\vo'|^{d+1-2i\nu}} \right]~.
\end{align}
Note that this is an $n \times n$ matrix with implicit spinor indices.

\subsection{Shadow transform}
The shadow transform \cite{Ferrara:1972xe, Ferrara:1972uq} is a non-local, linear transformation of conformal primary operators. In the scalar case, the transform is defined with a single integral on $\mathbb{R}^d$, but for fields with spin it involves also the uplift from $\mathbb{R}^d$ to the null cone of the embedding space \cite{Pasterski:2017kqt}. For (conjugate) Dirac spinors on $\mathbb{R}^d$, the calculation of the shadow transform proceeds in three steps \cite{Isono:2017grm}. First, uplift the field from $\vo \in \mathbb{R}^d$ to the null cone of the embedding space by writing 
\begin{equation}
\label{st:uplift}
	\Psi_\Delta(X;\vo) = \widehat{\Psi_\Delta}(X;q(\vo)) 
	\begin{pmatrix}
		\slo \\ 1	
	\end{pmatrix}
\end{equation}
and then dropping the dependence of $q$ on $\vo$. The uplifted $\widehat{\Psi_\Delta}(X;q)$ is defined modulo terms of the form $\Theta \Gamma^M q_M$, where $\Theta$ is an arbitrary (possibly $X$-dependent) spinor, because of \eqref{st:proj}.
However, such terms will drop out in the next step. Second, the shadow of the uplifted spinor is defined by\footnote{One may define the shadow with different normalization or phase factors.}
\begin{equation}
\label{st:shadow.uplift}
	 \widehat{\widetilde{\Psi_\Delta}}(X;q) = \int [\rmd q'] 
	 \frac{\widehat{\Psi_\Delta}(X;q') q'{}^M\Gamma_M}{(-2q\cdot q')^{d-\Delta+1/2}}~,
\end{equation}
where $[\rmd q']$ denotes the invariant integral measure for integration over the future light cone \cite{SimmonsDuffin:2012uy}.
In the last step, the shadow \eqref{st:shadow.uplift} is pulled back from the light-cone to $\mathbb{R}^d$ using again \eqref{st:uplift}. 
 
Let us implement this procedure. From \eqref{Dirac.sol}, we can easily recognize a representative of the uplift as 
\begin{equation}
\label{st:uplift.expl}
	\widehat{\Psi^\pm_\Delta}(X;q) = \int [\rmd \hp] \frac{\e{\pm i m \hp \cdot X}}{[-2 \hp\cdot q]^{\Delta+\frac12}} 
		\Pi_\pm(\hp)~.
\end{equation}
The calculation of the integral \eqref{st:shadow.uplift}, which we defer to App.~\ref{app.shadow}, results in
\begin{equation}
\label{st:st.uplift}
	 \widehat{\widetilde{\Psi^\pm_\Delta}}(X;q) =  \mp i \pi^{\frac{d}2}
	 \frac{\Gamma\left(\Delta - \frac{d-1}2 \right)}{\Gamma\left(\Delta + \frac12 \right)} 
	 \widehat{\Psi^\pm_{d-\Delta}}(X;q) + \Theta' \Gamma^M q_M~,
\end{equation}
where $\Theta'$ is a spinor which is known but not relevant for what follows. Indeed, applying \eqref{st:uplift} to pull back \eqref{st:st.uplift} from the light-cone to $\mathbb{R}^d$, the term containing $\Theta'$ drops out because of \eqref{st:proj}, and we obtain 
\begin{equation}
\label{st:st}
	 \widetilde{\Psi^\pm_\Delta}(X;\vo) = \mp i \pi^{\frac{d}2}
	 \frac{\Gamma\left(\Delta - \frac{d-1}2 \right)}{\Gamma\left(\Delta + \frac12 \right)} 
	 \Psi^\pm_{d-\Delta}(X;\vo)~.
\end{equation}

\subsection{Conformal primary basis}
Let us put together the results of the past two subsections. On the one hand, integrability of the Dirac inner product implies that the conformal dimensions of the wavefunctions \eqref{Dirac.sol} are restricted to the continuous principal series, $\Delta = \frac{d}2 +i \nu$, with $\nu\in \mathbb{R}$. On the other hand, the shadow transform \eqref{st:st}, which is a linear transformation, implies that the wavefunctions with negative $\nu$ are linear combinations of wavefunctions with positive $\nu$. Therefore, to get a basis of conformal primary, delta-function normalizable wavefunctions, we may restrict $\Delta$ to 
$\Delta = \frac{d}{2} + i\nu$ with $\nu\geq 0$. 

Henceforth, let us consider the \emph{normalized} wavefunctions
\begin{equation}
\label{sp:norm.wave}
	\Psi^\pm_{\frac{d}2 +i\nu}(X;\vo) = \left(\frac{m}{2\pi^2}\right)^{\frac{d+1}2} 
	\frac{\Gamma\left(\frac{d+1}2+i\nu\right)}{\Gamma\left(\frac12+i\nu\right)}
		\int [\rmd \hp] \frac{\e{\pm i m \hp \cdot X}}{[-2 \hp\cdot q(\vo)]^{\frac{d+1}2+i\nu}} 
		\Pi_\pm(\hp) \begin{pmatrix} \slo \\ 1 \end{pmatrix}~.
\end{equation}
Their shadow transforms are given by
\begin{equation}
\label{sp:st}
	 \widetilde{\Psi^\pm_{\frac{d}2 +i\nu}}(X;\vo) = \mp i \pi^{\frac{d}2}
	 \frac{\Gamma\left(\frac{1}2 -i\nu\right)}{\Gamma\left(\frac{d+1}2-i\nu \right)} 
	 \Psi^\pm_{\frac{d}2 -i\nu}(X;\vo)~.
\end{equation}
Restricting to $\nu> 0$, we have the following Dirac inner products,\footnote{The case $\nu=0$ is not included here.} 
\begin{align}
\label{sp:ip1}
	\left( \Psi^\pm_{\frac{d}2+i\nu}(\vo), \Psi^\pm_{\frac{d}2+i\nu'}(\vo')\right) &=
	\delta(\nu-\nu') \delta^d(\vo -\vo')~,\\
\label{sp:ip2}
	\left( \Psi^\pm_{\frac{d}2+i\nu}(\vo), \widetilde{\Psi^\pm_{\frac{d}2+i\nu'}}(\vo')\right) &=
	\delta(\nu-\nu') \frac{\slo -\slo'}{|\vo-\vo'|^{d+1-2i\nu}}~,\\
\label{sp:ip3}
	\left( \widetilde{\Psi^\pm_{\frac{d}2+i\nu}}(\vo), \Psi^\pm_{\frac{d}2+i\nu'}(\vo')\right) &=
	-\delta(\nu-\nu') \frac{\slo -\slo'}{|\vo-\vo'|^{d+1+2i\nu}}~,\\
\label{sp:ip4}
	\left( \widetilde{\Psi^\pm_{\frac{d}2+i\nu}}(\vo),  \widetilde{\Psi^\pm_{\frac{d}2+i\nu'}}(\vo')\right) &=
	\pi^d \frac{\left|\Gamma(\frac{1}2+i\nu)\right|^2}{\left|\Gamma(\frac{d+1}2+i\nu)\right|^2} \delta(\nu-\nu') \delta^d(\vo -\vo')~.
\end{align}
The third line can be obtained from the second one also from the identity $\left(\Psi, \Psi'\right) = \left(\Psi',\Psi\right)^\dagger$.

\subsection{Inverse transformation to the plane wave basis}

We are interested in the inverse of \eqref{sp:norm.wave}, \ie in expressing the plane-wave spinor wavefunctions in terms of the conformal primary basis. The main ingredient, which we prove in App.~\ref{app.complete}, is the completeness relation
\begin{equation}
\label{massive:complete}
	\frac1{\pi^{d+1}} \int\limits_0^\infty \rmd \nu 
	\frac{\left|\Gamma(\frac{d+1}2+i\nu)\right|^2}{\left|\Gamma(\frac12+i\nu)\right|^2} \int \rmd^d w\,
	\frac{\left[\Pi_\pm(\hp) \begin{pmatrix} \slo \\ 1 \end{pmatrix}\right]
	\overline{\left[\Pi_\pm(\hp') \begin{pmatrix} \slo \\ 1 \end{pmatrix}\right]}}{%
	[-2\hp \cdot q(\vo)]^{\frac{d+1}{2}+i\nu}[-2\hp' \cdot q(\vo)]^{\frac{d+1}{2}-i\nu}} 
	= \pm i\, \Pi_\pm(\hp)\, \delta^{d+1}(\hp,\hp')~,
\end{equation}
where $\delta^{d+1}(\hp,\hp')$ is the covariant delta function on $\mathbb{H}_{d+1}$.
Combining \eqref{massive:complete} with \eqref{sp:norm.wave}, one immediately finds
\begin{equation}
\label{massive:inverse}
	\Pi_\pm(\hp) \e{\pm im\hp\cdot X} = \mp i \left(\frac{2}m\right)^{\frac{d+1}2} 
	\int\limits_0^\infty \rmd \nu \int \rmd^d w\, \Psi^\pm_{\frac{d}2+i\nu}(X;\vo) 
	\frac{\Gamma(\frac{d+1}2-i\nu)}{\Gamma(\frac{1}2-i\nu)} 
	\frac{\overline{\left[\Pi_\pm(\hp) \begin{pmatrix} \slo \\ 1 \end{pmatrix}\right]}}{%
	[-2\hp \cdot q(\vo)]^{\frac{d+1}{2}-i\nu}}~.
\end{equation}
We remark that both formulas would also hold with integrals over negative $\nu$.

\section{Conformal primary basis for massless spinor fields}
\label{massless.case}

In this section, we will construct the conformal primary basis for massless fermions. To do so, we start with the normalized wavefunctions \eqref{sp:norm.wave} in the explicit form 
\begin{align}
\label{mf:norm.wave}
	\Psi^\pm_\Delta(X;\vo) &= \left(\frac{m}{2\pi^2}\right)^{\frac{d+1}2} 
	\frac{\Gamma\left(\Delta+\frac12\right)}{\Gamma\left(\Delta-\frac{d-1}2\right)}
		\int\limits_0^\infty \frac{\rmd y}{y^{d+1}} \int \rmd^d z \e{\pm i \left( \frac{m}y q(\vz) + my N_-\right) \cdot X}\\
\notag &\quad
		\times \left( \frac{y}{y^2 +|\vz -\vo|^2} \right)^{\Delta+\frac12}  
		\frac12 \left( 1 \pm \frac{i}{y} q^M(\vz) \Gamma_M \pm iy\Gamma_-\right) \begin{pmatrix} \slo \\ 1 \end{pmatrix}~.
\end{align}
Here, $N_-$ is the vector with components $N_-^M = \delta^M_-$ in light-cone coordinates. To obtain the behaviour for small $m$, we follow the method of \cite{Pasterski:2017kqt}. We replace the integration variable $y$ by $y \to m/y'$ and, keeping $y'$ fixed while taking $m$ small, make use of the distribution behaviour 
\begin{equation}
\label{mf:dist.limit}
	\left( \frac{y}{y^2 +|\vz -\vo|^2} \right)^{\Delta+\frac12} \underset{y\to 0}{\longrightarrow} 
	\pi^{\frac{d}2}\frac{\Gamma\left(\Delta-\frac{d-1}2\right)}{\Gamma\left(\Delta +\frac12 \right)}  y^{d-\Delta-\frac12} \delta^d(\vz-\vo) + \frac{y^{\Delta+\frac12}}{|\vz -\vo|^{2\Delta+1}} + \cdots~. 
\end{equation}
The ellipses indicate sub-leading terms in powers of $y^2$ with respect to both terms that are shown. 
Making use also of \eqref{st:proj}, this yields
\begin{equation}
\label{mf:wave.limit}
	\Psi^\pm_\Delta(X;\vo) \underset{m\ll 1}{\longrightarrow} m^{\frac{d}2-\Delta} \Upsilon^\pm_\Delta(X,\vo)
	 \pm m^{\Delta-\frac{d}2} 
	\frac{i\Gamma\left(\Delta+\frac12\right)}{\pi^{\frac{d}2}\Gamma\left(\Delta-\frac{d-1}2\right)} \int \rmd^d z\,
	\Upsilon^\pm_{d-\Delta}(X;\vz) \frac{\slo-\slz}{|\vo-\vz|^{2\Delta+1}} +\cdots~,
\end{equation}
where we have introduced the wavefunctions
\begin{equation}
\label{mf:massless.wavefunctions}
	\Upsilon^\pm_\Delta(X;\vo) = \frac{1}{\sqrt{2}(2\pi)^{\frac{d}2+1}} \int\limits_0^\infty \rmd y\, y^{\Delta-\frac12} 
	\e{\pm iy q(\vo)\cdot X} \begin{pmatrix} \slo \\ 1 \end{pmatrix}~.
\end{equation}

As in the case of scalar fields \cite{Pasterski:2017kqt}, there is no sensible $m\to 0$ limit of \eqref{mf:wave.limit}, but $\Upsilon^\pm_\Delta(X,\vo)$ are solutions of the massless Dirac equation and transform as conformal primaries under the conformal transformations induced on $\vo$ by the space-time Lorentz transformations. 
Furthermore, the integral in the second term of \eqref{mf:wave.limit} can be recognized as a shadow transform 
\begin{equation}
\label{mf:shadow}
	\widetilde{\Upsilon^\pm_\Delta}(X;\vo) = \int \rmd^d z\, \Upsilon^\pm_\Delta(X;\vz) \frac{\slo-\slz}{|\vo-\vz|^{2(d-\Delta)+1}}~.
\end{equation}  

The Dirac inner product of two wavefunctions \eqref{mf:massless.wavefunctions} is straightforward. As in the massive case, integrability restricts $\Delta$ to the principal continuous series, $\Delta \in \frac{d}2 +i\mathbb{R}$. 
The result is 
\begin{equation}
\label{mf:ip}
	\left( \Upsilon^\pm_{\frac{d}2+i\nu}(\vo), \Upsilon^\pm_{\frac{d}2+i\nu'}(\vo') \right) = \delta(\nu-\nu') \delta^d(\vo-\vo')
\end{equation}
for any $\nu, \nu'$. 

It remains to express the plane wave solutions in terms of the conformal primary fields. This is entirely equivalent to the scalar case \cite{Pasterski:2017kqt}, because we recognize a Mellin transform in \eqref{mf:massless.wavefunctions},
\begin{equation}
\label{mf:Mellin}
	\int\limits_0^\infty \rmd y \,y^{\Delta-\frac12} 
	\e{\pm i y q(\vo)\cdot X - \epsilon y}
	= \frac{(\mp i)^{\Delta+\frac12} \Gamma(\Delta+\frac12)}{[- q(\vo)\cdot X \mp i \epsilon]^{\Delta+\frac12}}~.
\end{equation}
Inverting it, one obtains
\begin{equation}
\label{mf:inverse.Mellin}
	\e{\pm i y q(\vo)\cdot X} 
	\begin{pmatrix}
		\slo \\1
	\end{pmatrix} =
	\sqrt{2} (2\pi)^{\frac{d}2} \int\limits_{-\infty}^{\infty} \rmd \nu \,y^{-\frac{d+1}2-i\nu} \,
	\Upsilon^\pm_{\frac{d}2+i\nu}(X;\vo)~.
\end{equation}

Therefore, in the massless case, the wavefunctions \eqref{mf:massless.wavefunctions} with $\Delta \in \frac{d}2 +i\mathbb{R}$ form a complete basis of conformal primary spinor wavefunctions.

%\input{concs}

%%%%%%%%%%%%%%%%%%%%%%%%%%%%%%%%%%%%%%%%%%%%
%\input{results}

%\input{completeness}
%%%%%%%%%%%%%%%%%%%%%%%%%%%%%%%%%%%%%%%%%%%%
%\input{Concs}
%%%%%%%%%%%%%%%%%%%%%%%%%%%%%%%%%%%%%%%%%%%%
%
\section*{Acknowledgements}
We are grateful to Massimo Taronna for helpful suggestions and comments on the maniscript.
This work was supported partly by the INFN, research initiative STEFI.
%%%%%%%%%%%%%%%%%%%%%%%%%%%%%%%%%%%%%%%%%%%%

\begin{appendix}
\section{Recurring formulae}
\label{app:formulae}

In this appendix, we collect a few formulae, which will recur in the detailed calculations in the appendices that follow.

\emph{Schwinger parameterization} 
\begin{equation}
\label{appf:Schwinger}
	\frac1{A^a} = \frac1{\Gamma(a)} \int\limits_0^\infty \rmd s\, s^{a-1} \e{-sA}~. 
\end{equation}

Provided that $\re a>0$ and $\re b>0$, the following two-factor numerator can be rewritten using \emph{Feynman parameters},
\begin{equation}
\label{appf:Feynm}
	\frac{1}{A^a B^b} = \frac{\Gamma(a+b)}{\Gamma(a)\Gamma(b)} \int\limits_0^1 \rmd \alpha \int\limits_0^1 \rmd \beta\,
	 \delta(\alpha+\beta-1)\, \frac{\alpha^{a-1}\beta^{b-1}}{(\alpha A +\beta B)^{a+b}}~.
\end{equation}

For time-like $X$ ($X^2<0$), we have the following integral on the projective light cone \cite{SimmonsDuffin:2012uy},
\begin{equation}
\label{appf:conf.int}
	\int [\rmd q] \frac1{(-2X\cdot q)^d} = \int \rmd^d w\, \frac1{[-2X\cdot q(\vo)]^d} 
	= \frac{\pi^{\frac{d}{2}} \Gamma(\frac{d}{2})}{\Gamma(d)}\left(-X^2\right)^{-\frac{d}2}~.
\end{equation}

\section{Calculation of the explicit form}
\label{app.explicit}

Here, we obtain the explicit form of the conformal primary wavefunctions by performing a direct evaluation of the momentum integral in \eqref{Dirac.sol}. We start by rewriting it in the form
\begin{equation}
\label{Dirac.sol.Phi}
	\Psi^\pm_\Delta(X;\vo) = \frac12 \left( 1+\frac1m \Gamma^M \partial_M \right)
	  \Phi^\pm_{\Delta+\frac12} (X;\vo) \begin{pmatrix} \slo \\ 1 \end{pmatrix}~,
\end{equation}
where the $\Phi^\pm_{\Delta+\frac12}(X;\vo)$ is a scalar primary wavefunction,
\begin{equation}
\label{Phi.primary}
	\Phi^\pm_{\Delta} (X;\vo) = \int [\rmd \hp] 
	\frac{\e{\pm i m \hp \cdot X}}{[-2 \hp\cdot q(\vo)]^{\Delta}}~.
\end{equation}
Its explicit form can be found in \cite{Pasterski:2017kqt}, but we shall include the full calculation here for comprehensiveness. 

To evaluate \eqref{Phi.primary}, we parameterize $\hp$ in terms of $y$ and $\vz$ using \eqref{emb:H.coord}, 
\begin{equation}
\label{Phi.primary.2}
	\Phi^\pm_{\Delta} (X;\vo) = \int \frac{\rmd y \rmd^d z}{y^{d+1}} 
	\left( \frac{y}{y^2+|\vz -\vo|^2}\right)^\Delta  
	\e{\pm i m \hp \cdot X}~.
\end{equation}
Shifting $\vz \to \vz+\vo$ results in
\[ \hp \cdot X \to \frac1y \left[ q(\vo)\cdot X - \frac12 X^+ (y^2 + \vz^2) - \vz \cdot (X^+\vo -\vec{X})\right]~, \]
and we can evaluate the angular part of the $\vz$-integral in \eqref{Phi.primary.2}. This yields
\begin{align}
\label{Phi.primary.3}
	\Phi^\pm_{\Delta} (X;\vo) &= \int \frac{\rmd y \rmd z}{y^{d+1}} 
	\left( \frac{y}{y^2+z^2}\right)^\Delta (2\pi z)^{\frac{d}2} 
	\left( \frac{m}y |X^+\vo -\vec{X}| \right)^{1-\frac{d}2} \\
\notag 
 &\quad \times
	\e{\pm i \frac{m}y \left[q(\vo) \cdot X - \frac12 X^+ (z^2+y^2)\right]}
	\BesselJ_{\frac{d}2-1}\left( \frac{zm}y|X^+\vo -\vec{X}|\right)~,
\end{align}
with $\BesselJ_\alpha$ denoting a Bessel function.
Notice that $|X^+\vo -\vec{X}|^2 = X^2 - 2X^+ q(\vo)\cdot X$.
Now, rescaling $z \to yz$ results in 
\begin{align}
\label{Phi.primary.4}
	\Phi^\pm_{\Delta} (X;\vo) &= (2\pi)^{\frac{d}2} 
	\left( m |X^+\vo -\vec{X}| \right)^{1-\frac{d}2}
	\int\limits_0^\infty \frac{\rmd y}y y^{-\Delta} 
	\int\limits_0^\infty \rmd z \frac{z^{\frac{d}2}}{(1+z^2)^\Delta}
	\\
\notag 
 &\quad \times
	\e{\pm i \frac{m}y \left[q(\vo) \cdot X-\frac12 X^+ (z^2+1)y^2\right]}
	\BesselJ_{\frac{d}2-1}\left(zm|X^+\vo -\vec{X}|\right)~.
\end{align}
We proceed by performing Schwinger parameterization of $(1+z^2)^{-\Delta}$,
after which we can carry out the $z$-integral in \eqref{Phi.primary.4}. This results in
\begin{align}
\label{Phi.primary.5}
	\Phi^\pm_{\Delta} (X;\vo) &= \frac{(2\pi)^{\frac{d}2}}{\Gamma(\Delta)} 
	\int\limits_0^\infty \frac{\rmd y}y
	\int\limits_0^\infty \frac{\rmd s}s
	\left(\frac{s}{y}\right)^{\Delta} 
	\left( 2s \pm im yX^+ \right)^{-\frac{d}2}
	\\
\notag 
 &\quad \times
	\exp\left[-\left(s \pm \frac12 i m y X^+\right) 
	\pm i \frac{m}y q(\vo)\cdot X
	-\frac{m^2 (X^2 - 2X^+ q(\vo)\cdot X)}{2(2s\pm imyX^+)}\right]~.
\end{align}
Other two rescalings, first $s \to sy$ and then $y \to y(s\pm \frac12 im X^+)^{-1}$,\footnote{This assumes that the order of the two integrations can be safely interchanged, which is justified after including the regulator in the $s$-integral.} result in
\begin{equation}
\label{Phi.primary.6}
	\Phi^\pm_{\Delta} (X;\vo) = \frac{\pi^{\frac{d}2}}{\Gamma(\Delta)} 
	\int\limits_0^\infty \frac{\rmd y}y y^{-\frac{d}2}
	\e{-y -\frac{m^2 X^2}{4y}}
	\int\limits_0^\infty \frac{\rmd s}s s^\Delta 
	\e{- \frac{m}{y} \left[\epsilon \mp i q(\vo)\cdot X \right] s}~.
\end{equation}
Here, we have inserted a regulator in the exponent of the integrand of the $s$-integral in order to make it convergent. 
The result of the integration is
\begin{align}
\notag 
	\Phi^\pm_{\Delta} (X;\vo) &= 
	\frac{\pi^{\frac{d}2}}{[\pm im(-q(\vo)\cdot X\mp i\epsilon)]^\Delta}
	\int\limits_0^\infty \frac{\rmd y}y y^{\Delta-\frac{d}2}
	\e{-y -\frac{m^2 X^2}{4y}} \\
\label{Phi.primary.7}
  &=  \frac{2(2\pi)^{\frac{d}2} (\pm im)^{-\Delta}}{[-2q(\vo)\cdot X\mp i\epsilon]^\Delta}
  \left(m\sqrt{X^2}\right)^{\Delta-\frac{d}2} 
  \BesselK_{\Delta-\frac{d}2} \left(m\sqrt{X^2}\right)~.
\end{align}
Finally, substituting \eqref{Phi.primary.7} into \eqref{Dirac.sol.Phi} yields
\eqref{Dirac.sol.explicit} in the main text.

\section{Calculation of the Dirac inner product}
\label{app.dirac}

In this appendix, we will present the detailed calculation of the Dirac inner product of two wavefunctions of the form \eqref{Dirac.sol}. 
To simplify the calculation, observe that the inner product \eqref{inner.prod} is invariant under Lorentz transformations, and they act as conformal transformations on $\vo$. Thus, it is straightforward to show that
\begin{equation}
\label{ip:invariance}
	\left( \Psi^\pm_\Delta(\vo), \Psi^\pm_{\Delta'}(\vo') \right) = \left( \Psi^\pm_\Delta(0), \Psi^\pm_{\Delta'}(\vo'-\vo) \right)~.
\end{equation}
Therefore, without loss of generality, we will consider 
\begin{align}
\label{ip:start}
	 \left( \Psi^\pm_\Delta(0), \Psi^\pm_{\Delta'}(\vo') \right) = \int \rmd^{d+1}X \int [\rmd \hp][\rmd \hp'] \e{\pm im(\hp'-\hp)\cdot X}
	 \frac{(0,1)\Pi_\pm^\dagger(\hp)\Pi_\pm(\hp')(\slo', 1)^T}{[-2\hp\cdot q(0)]^{\Delta^\ast+\frac12} [-2\hp'\cdot q(\vo')]^{\Delta'+\frac12}}~.
\end{align}
Performing first the integrals over $X$ and $\hp'$ yields 
\begin{align}
\label{ip:1}
	 \left( \Psi^\pm_\Delta(0), \Psi^\pm_{\Delta'}(\vo') \right) = \left(\frac{2\pi}m\right)^{d+1} \int [\rmd \hp] \frac1{\hp^0} 
	 \frac{(0,1)\Pi_\pm^\dagger(\hp)\Pi_\pm(\hp)(\slo', 1)^T}{[-2\hp\cdot q(0)]^{\Delta^\ast+\frac12} [-2\hp\cdot q(\vo')]^{\Delta'+\frac12}}~.
\end{align}
We proceed with some algebra in the numerator, for which we find
\begin{equation}
\label{ip:num.algebra} 
	\Pi_\pm^\dagger(\hp)\Pi_\pm(\hp) =\pm i \hp^0 \Gamma_0  \Pi_\pm(\hp)~,
\end{equation}
so that \eqref{ip:1} becomes
\begin{equation}
\label{ip:2}
	 \left( \Psi^\pm_\Delta(0), \Psi^\pm_{\Delta'}(\vo') \right) = \frac12 \left(\frac{2\pi}m\right)^{d+1} \int [\rmd \hp]  
	 \frac{\hp^- - \slashed{\hp} \slo' \pm i \slo'}{[-2\hp\cdot q(0)]^{\Delta^\ast+\frac12} [-2\hp\cdot q(\vo')]^{\Delta'+\frac12}}~.
\end{equation}
Note that $\slashed{\hp}=\hp^i \gamma_i$ involves only the $d$ Euclidean components. 
Next, we write $\hp$ in terms of the $\mathbb{H}_{d+1}$-coordinates \eqref{emb:H.coord}, with the integral measure given in \eqref{emb:H.integ}. Moreover, we introduce $\vz'=\vo'-\vz$, which is fixed using a delta function which, in turn, is written as a plane wave expansion. This transforms \eqref{ip:2} into 
\begin{align}
\label{ip:3}
	 \left( \Psi^\pm_\Delta(0), \Psi^\pm_{\Delta'}(\vo') \right) &= \frac12 \left(\frac{2\pi}m\right)^{d+1} \int \frac{\rmd^d k}{(2\pi)^d} 
	 \e{-i \vk\cdot \vo'} 
	 \int\limits_0^\infty \frac{\rmd y}{y} y^{\Delta^\ast + \Delta'-d}\\
\notag 
	&\quad \times \int \rmd^d z \rmd^d z' \e{i \vk\cdot(\vz +\vz')} 
	\frac{ y^2 \pm iy (\slz +\slz') - \slz\slz'}{(y^2 +|\vz|^2)^{\Delta^\ast+\frac12} (y^2 +|\vz'|^2)^{\Delta'+\frac12}}~.
\end{align}
Another rescaling, $\vz\to y \vz$ and $\vz'\to y\vz'$, gives rise to
\begin{align}
\label{ip:4}
	 \left( \Psi^\pm_\Delta(0), \Psi^\pm_{\Delta'}(\vo') \right) &= \frac12 \left(\frac{2\pi}m\right)^{d+1} \int \frac{\rmd^d k}{(2\pi)^d} 
	 \e{-i \vk\cdot \vo'} 
	 \int\limits_0^\infty \frac{\rmd y}{y} y^{-\Delta^\ast - \Delta'+d}\\
\notag 
	&\quad \times \int \rmd^d z \e{i y\vk\cdot\vz} \frac{1\pm i \slz}{(1+|\vz|^2)^{\Delta^\ast+\frac12}}
	\int \rmd^d z' \e{i y\vk\cdot\vz'} \frac{1\pm i \slz'}{(1+|\vz'|^2)^{\Delta'+\frac12}}~.
\end{align}

Now, we need to deal with the two integrals on the second line of \eqref{ip:4}, which are identical in form. Let us consider the first one. We pull out the numerator, replacing $\vz$ by a derivative with respect to $\vk$, and use Schwinger parameterization \eqref{appf:Schwinger} for the denominator. This yields 
\begin{equation}
\label{ip:zint.1}
	\int \rmd^d z \e{i y \vk\cdot\vz} \frac{1\pm i \slz}{(1+|\vz|^2)^{\Delta^\ast+\frac12}} = 
	\left( 1\pm \frac1y \gamma^i \frac{\partial}{\partial k^i} \right) \int \rmd^d z \e{i y \vk\cdot\vz} 
	\int\limits_0^\infty \rmd s\, \frac{s^{\Delta^\ast-\frac12}}{\Gamma(\Delta^\ast+\frac12)} \e{-s(1+|\vz|^2)}
\end{equation}   
After exchanging the order of integration and completing the square in the exponent, the integral over $\vz$ is Gaussian and results in
\begin{equation}
\label{ip:zint.2}
	\int \rmd^d z \e{i y\vk\cdot\vz} \frac{1\pm i \slz}{(1+|\vz|^2)^{\Delta^\ast+\frac12}} = 
	\frac{\pi^{\frac{d}{2}}}{\Gamma(\Delta^\ast+\frac12)}  
	\int \limits_0^\infty \rmd s\, s^{\Delta^\ast-\frac{d+1}2}\left(1\mp \frac{y}{2s} \slk\right) \e{-s -\frac{y^2k^2}{4s}}~.
\end{equation} 

Returning to \eqref{ip:4}, we substitute \eqref{ip:zint.2} and the analogous result for the other integral and proceed by rescaling $y\to 2y/k$, with $k=|\vk|$. This yields 
\begin{align}
\label{ip:5}
	 \left( \Psi^\pm_\Delta(0), \Psi^\pm_{\Delta'}(\vo') \right) &= \frac12 \left(\frac{2\pi}m\right)^{d+1} 
	 \frac{\pi^d}{\Gamma(\Delta^\ast+\frac12)\Gamma(\Delta'+\frac12)}  
	 \int \frac{\rmd^d k}{(2\pi)^d} \e{-i \vk\cdot \vo'} 
	 \int\limits_0^\infty \frac{\rmd y}{y} \left(\frac{2y}k\right)^{-\Delta^\ast-\Delta'+d} \\
\notag 
	&\quad \times \int \limits_0^\infty \rmd s\, s^{\Delta^\ast-\frac{d+1}2}\left(1 \mp \frac{y}{sk} \slk \right) \e{-s -\frac{y^2}{s}}
	\int \limits_0^\infty \rmd t\, t^{\Delta'-\frac{d+1}2}\left(1 \mp \frac{y}{tk} \slk \right) \e{-t -\frac{y^2}{t}}~.
\end{align}
Another rescaling, $s\to ys$ and $t\to yt$, renders the $y$-integral elementary and gives
\begin{align}
\label{ip:6}
	 \left( \Psi^\pm_\Delta(0), \Psi^\pm_{\Delta'}(\vo') \right) &= \frac12 \left(\frac{2\pi}m\right)^{d+1} 
	 \frac{\pi^d}{\Gamma(\Delta^\ast+\frac12)\Gamma(\Delta'+\frac12)}  
	 \int \frac{\rmd^d k}{(2\pi)^d} \e{-i \vk\cdot \vo'} \left(\frac{k}2\right)^{\Delta^\ast+\Delta'-d}
	 \\
\notag 
	&\quad \times \int \limits_0^\infty \rmd s\, \int \limits_0^\infty \rmd t\,
	\frac{s^{\Delta^\ast-\frac{d+1}2}t^{\Delta'-\frac{d+1}2}}{s+\frac1s+t+\frac1t} 
	\left(1\mp\frac{1}{sk} \slk \right) \left(1\mp\frac{1}{tk} \slk \right)~.
\end{align}
In order to deal with the double integral on the second line of \eqref{ip:6}, we change variables by setting $s=\e{U+V}$ and $t=\e{U-V}$,
\begin{multline}
\label{ip:st.int}
	\int \limits_0^\infty \rmd s\, \int \limits_0^\infty \rmd t\,
	\frac{s^{\Delta^\ast-\frac{d+1}2}t^{\Delta'-\frac{d+1}2}}{s+\frac1s+t+\frac1t} 
	\left(1 \mp \frac{1}{sk} \slk \right) \left(1 \mp \frac{1}{tk} \slk \right)
	\\ = 
	\int\limits_{-\infty}^\infty \rmd U \int\limits_{-\infty}^\infty \rmd V\, \e{U(\Delta^\ast+\Delta'-d)} \e{V(\Delta^\ast-\Delta')}
	\left( \frac1{\cosh V} \mp \frac{\frac1k \slk}{\cosh U} \right)~.
\end{multline}
At this point, the integrals in $U$ and $V$, in order to be defined, impose that the conformal dimensions must be
\begin{equation}
\label{ip:Delta}
	\Delta = \frac{d}2 +i \nu~,\qquad \nu \in \mathbb{R}~.
\end{equation}
Then, the result of integration in \eqref{ip:st.int} is
\begin{multline}
\label{ip:UV.int}
	\int \limits_0^\infty \rmd s\, \int \limits_0^\infty \rmd t\,
	\frac{s^{\Delta^\ast-\frac{d+1}2}t^{\Delta'-\frac{d+1}2}}{s+\frac1s+t+\frac1t} 
	\left(1 \mp \frac{1}{sk} \slk \right) \left(1 \mp \frac{1}{tk} \slk \right)
	\\ = 
	2\pi \Gamma\left(\frac12+i\nu\right)\Gamma\left(\frac12-i\nu\right) \left[ \delta(\nu -\nu') \mp \frac{1}{k} \slk \delta(\nu +\nu') \right]~.
\end{multline}
After inserting \eqref{ip:UV.int} into \eqref{ip:6} and performing the remaining integral over $\vk$, we obtain the final result
\begin{align}
\label{ip:7}
	 \left( \Psi^\pm_{\frac{d}2+i\nu}(0), \Psi^\pm_{\frac{d}2+i\nu'}(\vo') \right) &= \left(\frac{2\pi^2}m\right)^{d+1} 
	 	\left[ \frac{\Gamma(\frac12-i\nu) \Gamma(\frac12+i\nu)}{\Gamma(\frac{d+1}2 -i\nu) \Gamma(\frac{d+1}2 +i\nu)}
	 	\delta(\nu - \nu') \delta^d(\vo') \right.\\
\notag
	&\quad \left. \mp i \frac{\Gamma(\frac12-i\nu)}{\pi^{d/2} \Gamma(\frac{d+1}2 -i\nu)}\delta(\nu + \nu') 
	\frac{\slo'}{|\vo'|^{d+1-2i\nu}} \right]~.
\end{align}

\section{Calculation of the shadow transform}
\label{app.shadow}

In this appendix, we calculate the shadow transform of the uplift
\begin{align}
\notag
	\widehat{\widetilde{\Psi^\pm_\Delta}}(X;q) &= \int [\rmd q'] 
	\frac{\widehat{\Psi^\pm_\Delta}(X;q') \Gamma^M q'_M}{[-2q\cdot q']^{d-\Delta+\frac12}}\\
\notag
	&=	\int [\rmd q']  \int [\rmd \hp] 
	\frac{\e{\pm im\hp\cdot X} \Pi_\pm(\hp) \Gamma^M q'_M}{[-2\hp \cdot q']^{\Delta+\frac12}[-2q\cdot q']^{d-\Delta+\frac12}}\\
\label{as:1}
	&= \int [\rmd \hp] \frac{\e{\pm im\hp\cdot X} \Pi_\pm(\hp)}{2\Delta-1} \Gamma^M\frac{\partial}{\partial\hp^M} 
	\int [\rmd q'] \frac{1}{[-2\hp \cdot q']^{\Delta-\frac12}[-2q\cdot q']^{d-\Delta+\frac12}}~. 
\end{align}
Using Feynman parameterization \eqref{appf:Feynm} for the two-term denominator renders the $q'$-integral in the form \eqref{appf:conf.int}. After performing that integral, \eqref{as:1} becomes
\begin{align}
\label{as:2}
	\widehat{\widetilde{\Psi^\pm_\Delta}}(X;q) &= 
	\frac{\pi^{\frac{d}2} \Gamma(\frac{d}2+1)}{\Gamma(\Delta+\frac12) \Gamma(d-\Delta+\frac12)}
	\int [\rmd \hp] \e{\pm im\hp\cdot X} \Pi_\pm(\hp) \\
\notag &\quad \times \int\limits_0^1 \frac{\rmd\alpha}{\alpha(1-\alpha)} 
	\left(\frac{\alpha}{1-\alpha}\right)^{\Delta+\frac12} 
	\frac{\Gamma^M(\frac{\alpha}{1-\alpha} \hp + q)_M}{[-(\frac{\alpha}{1-\alpha} \hp +q)^2]^{\frac{d}2+1}}.
\end{align}
We remark that one must use $\hp^2=-1$ only after performing the differentiation with respect to $\hp^M$. Next, we change
integration variable by setting $\frac{\alpha}{1-\alpha}=(-2\hp \cdot q) z$, which gives
\begin{align}
\label{as:3}
	\widehat{\widetilde{\Psi^\pm_\Delta}}(X,q) &= 
	\frac{\pi^{\frac{d}2} \Gamma(\frac{d}2+1)}{\Gamma(\Delta+\frac12) \Gamma(d-\Delta+\frac12)}
	\int [\rmd \hp] \frac{\e{\pm im\hp\cdot X} \Pi_\pm(\hp)}{[-2\hp\cdot q]^{d-\Delta+\frac12}} \\
\notag &\quad \times \int\limits_0^\infty \frac{\rmd z}{z} 
	\left[ \Gamma^M \hp_M \frac{z^{\Delta-\frac{d-1}2}}{(1+z)^{\frac{d}2+1}} 
	+ \frac1{[-2\hp\cdot q]} \Gamma^M q_M \frac{z^{\Delta-\frac{d+1}2}}{(1+z)^{\frac{d}2+1}} \right]~. 
\end{align}
Finally, carrying out the $z$-integral results in
\begin{equation}
\label{as:4}
	 \widehat{\widetilde{\Psi^\pm_\Delta}}(X;q) =  \mp i \pi^{\frac{d}2}
	 \frac{\Gamma\left(\Delta - \frac{d-1}2 \right)}{\Gamma\left(\Delta + \frac12 \right)} 
	 \int [\rmd \hp] \frac{\e{\pm i\hp \cdot X}\Pi_\pm (\hp)}{[-2\hp \cdot q]^{d-\Delta+\frac12}}
	 \left[ 1 \pm i \frac{d-\Delta+\frac12}{\Delta-\frac{d+1}2} \frac{\Gamma^M q_M}{[-2\hp \cdot q]} 
	 \right]~.
\end{equation}
\section{Proof of the completeness relation}
\label{app.complete}

Here, we shall calculate the integral
\begin{equation}
\label{app.comp:I.def.3}
	\mathcal{I}(\hp, \hp') = \frac1{\pi^{d+1}} \int\limits_0^\infty \rmd \nu 
	\frac{\left|\Gamma(\frac{d+1}2+i\nu)\right|^2}{\left|\Gamma(\frac12+i\nu)\right|^2} \int \rmd^d w\,
	\frac{\left[\Pi_\pm(\hp) \begin{pmatrix} \slo \\ 1 \end{pmatrix}\right]
	\overline{\left[\Pi_\pm(\hp') \begin{pmatrix} \slo \\ 1 \end{pmatrix}\right]}}{%
	[-2\hp \cdot q(\vo)]^{\frac{d+1}{2}+i\nu}[-2\hp' \cdot q(\vo)]^{\frac{d+1}{2}-i\nu}}~.
\end{equation}
First, manipulate the numerator in the integral,
\begin{equation}
\label{app.comp:num.alg}
	\left[\Pi_\pm(\hp) \begin{pmatrix} \slo \\ 1 \end{pmatrix}\right]
	\overline{\left[\Pi_\pm(\hp') \begin{pmatrix} \slo \\ 1 \end{pmatrix}\right]}
	= - \Pi_\pm(\hp) \Gamma^M q_M(\vo) \Pi_\pm(\hp')~,
\end{equation}
so that
\begin{equation}
\label{app.comp:I.7}
	\mathcal{I}(\hp, \hp') = -\frac1{\pi^{d+1}} \int\limits_0^\infty \rmd \nu 
	\frac{\left|\Gamma(\frac{d+1}2+i\nu)\right|^2}{\left|\Gamma(\frac12+i\nu)\right|^2} \int \rmd^d w\,
	\frac{\Pi_\pm(\hp) \Gamma^M q_M(\vo) \Pi_\pm(\hp')}{%
	[-2\hp \cdot q(\vo)]^{\frac{d+1}{2}+i\nu}[-2\hp' \cdot q(\vo)]^{\frac{d+1}{2}-i\nu}}~.
\end{equation}
Next, consider the integral in $\vo$
\begin{align}
\notag
	I_{\vo} &= 
	\int \rmd^d w\,
	\frac{\Gamma^M q_M(\vo)}{%
	[-2\hp \cdot q(\vo)]^{\frac{d+1}{2}+i\nu}[-2\hp' \cdot q(\vo)]^{\frac{d+1}{2}-i\nu}} \\
\notag 
	&= \frac{\Gamma^M}{2(\frac{d-1}{2}+i\nu)} \frac{\partial}{\partial \hp^M} \int \rmd^d w\,
	\frac{1}{%
	[-2\hp \cdot q(\vo)]^{\frac{d-1}{2}+i\nu}[-2\hp' \cdot q(\vo)]^{\frac{d+1}{2}-i\nu}} \\
\label{app.comp:w.int.spin}
	&= \frac{\Gamma(d)}{2 \left|\Gamma(\frac{d+1}2+i\nu)\right|^2} \Gamma^M  \frac{\partial}{\partial \hp^M}
	\int \rmd^d w\, \int\limits_0^1 \frac{\rmd \alpha}{\alpha(1-\alpha)} 
	\frac{\alpha^{\frac{d-1}2+i\nu}(1-\alpha)^{\frac{d+1}2-i\nu}}{\{-2[\alpha \hp+(1-\alpha) \hp'] \cdot q(\vo)\}^{d}}~,
\end{align}
where we have used Feynman parameterization \eqref{appf:Feynm} for the denominator of the integrand. To be able to apply \eqref{appf:conf.int}, we need to check that the vector in the square bracket in the denominator is time-like. We emphasize that we cannot set $\hp^2=-1$ yet, because of the formal differentation. Nevertheless,
\begin{equation}
	[\alpha \hp+(1-\alpha) \hp']^2 = -1 -\alpha(1-\alpha)(\hp-\hp')^2 +\alpha (\hp^2+1)<0~,
\end{equation}
because the last summand can be considered as infinitesimal. Therefore,
\begin{align}
\notag
	I_{\vo} &= \frac{\pi^{\frac{d}2}\Gamma(\frac{d}2)}{2 \left|\Gamma(\frac{d+1}2+i\nu)\right|^2}
	\Gamma^M  \frac{\partial}{\partial \hp^M} \int\limits_0^1 \frac{\rmd \alpha}{\alpha(1-\alpha)} 
	\frac{\alpha^{\frac{d-1}2+i\nu}(1-\alpha)^{\frac{d+1}2-i\nu}}{%
	[1 + \alpha(1-\alpha)(\hp-\hp')^2 - \alpha (\hp^2+1)]^{\frac{d}2}}\\
\label{app.comp:w.int.spin.2}
	&= \frac{\pi^{\frac{d}2}\Gamma(\frac{d}2+1)}{\left|\Gamma(\frac{d+1}2+i\nu)\right|^2}
	\int\limits_0^1 \frac{\rmd \alpha}{\alpha(1-\alpha)} \frac{\alpha^{\frac{d+1}2+i\nu}(1-\alpha)^{\frac{d+1}2-i\nu}\,%
	\Gamma^M [\hp - (1-\alpha) (\hp-\hp')]_M}{[1 + \alpha(1-\alpha)(\hp-\hp')^2]^{\frac{d}2+1}}~,
\end{align}
where we have set $\hp^2=-1$ after the differentiation. 

Returning to \eqref{app.comp:I.7}, where $I_{\vo}$ appears sandwiched between the two projectors $\Pi_\pm(\hp)$ and $\Pi_\pm(\hp')$, we observe that the term with $(\hp-\hp')$ in the numerator of \eqref{app.comp:w.int.spin.2} does not contribute. The remaining term, after changing the integration variable by setting $\e{u}= \frac{\alpha}{1-\alpha}$, gives rise to
\begin{equation}
\label{app.comp:I.8}
 	\mathcal{I}(\hp, \hp') = \pm  \frac{i \Gamma(\frac{d}2+1)}{\pi^{\frac{d}2+2}} \Pi_\pm(\hp) \Pi_\pm(\hp')
 	\int\limits_0^\infty \rmd \nu\, \cosh(\pi\nu) \int\limits_{-\infty}^{\infty}\rmd u\,
    \frac{2\cosh \frac{u}2\, \e{iu\nu}}{[2+2\cosh u + (\hp-\hp')^2]^{\frac{d}2+1}}~.
\end{equation} 
Exploiting the symmetries of the integrand, this can be rewritten as
\begin{equation}
\label{app.comp:I.81}
 	\mathcal{I}(\hp, \hp') = \pm  \frac{i \Gamma(\frac{d}2+1)}{2\pi^{\frac{d}2+2}} \Pi_\pm(\hp) \Pi_\pm(\hp')
 	\int\limits_{-\infty}^\infty \rmd \nu\, \cosh(\pi\nu) \int\limits_{-\infty}^{\infty}\rmd u\,
    \frac{2\cosh \frac{u}2\, \e{iu\nu}}{[2+2\cosh u + (\hp-\hp')^2]^{\frac{d}2+1}}~.
\end{equation} 
Now, in order to be able to exchange the order of integration, we need to regulate the $\nu$-integral by introducing a convergence factor $\e{-\varepsilon \nu^2}$. With 
\begin{equation}
\label{app.comp:nu.int.2}
	\int\limits_{-\infty}^\infty \rmd \nu\, \e{-\varepsilon \nu^2} \cosh(\pi\nu) \e{iu\nu} 
	= \frac1{2} \sqrt{\frac{\pi}{\varepsilon}}  
	\left[ \e{-\frac{(u-i\pi)^2}{4\varepsilon}} + \e{-\frac{(u+i\pi)^2}{4\varepsilon}} \right]
\end{equation}
and using the symmetry of the integrand of the $u$-integral, we get 
\begin{equation}
\label{app.comp:I.9}
 	\mathcal{I}(\hp, \hp') = \pm  \frac{i \Gamma(\frac{d}2+1)}{2\pi^{\frac{d+3}2}\sqrt{\varepsilon}} \Pi_\pm(\hp) \Pi_\pm(\hp')
 	\int\limits_{-\infty}^{\infty} \rmd u\,
    \frac{2\cosh \frac{u}2\, \e{-\frac{(u-i\pi)^2}{4\varepsilon}}}{[2+2\cosh u + (\hp-\hp')^2]^{\frac{d}2+1}}~.
\end{equation} 
Considering this integral in the complex $u$-plane, we may deform the integration contour such that it passes very close to the point $u=i\pi$. Then, for $\varepsilon\to0$, the support of the integrand is restricted to the neighbourhood of this point. For $u=i(\pi-\delta)$, $|\delta|\ll 1$, we have
\begin{equation}
\label{app.comp:integrand.u.ipi.2}
	\frac{2 \cosh \frac{u}2}{ [2+2\cosh u + (\hp-\hp')^2]^{\frac{d}2+1}} \approx \frac{\delta}{ [\delta^2 + (\hp-\hp')^2]^{\frac{d}2+1}} 
	\underset{\delta\ll 1}{\longrightarrow}  \frac{\pi^{\frac{d}2+1}}{\Gamma(\frac{d}2+1)} \delta^{d+1}(\hp,\hp')~,  
\end{equation}
where we have refrained from writing subleading terms in $\delta$. 
Thus, after performing the remaining Gaussian integral, the final result is 
\begin{equation}
\label{app.comp:I.10}
	\mathcal{I}(\hp, \hp') = \pm i\, \Pi_\pm(\hp)\, \delta^{d+1}(\hp,\hp')~.
\end{equation}
\end{appendix}

%%%%%%%%%%%%%%%%%%%%%%%%%%%%%%%%%%%%%%%%%%%%
\bibliographystyle{JHEPnotes}
\bibliography{assympt-sym}
\end{document}